\documentclass[12pt]{article}
\usepackage{amsmath,amssymb,graphicx} 
\def\be{\begin{equation}}
\def\ee{\end{equation}}
\def\bea{\begin{eqnarray}}
\def\eea{\end{eqnarray}}
\def\a{\alpha}
\def\b{\beta}
\def\g{\gamma}
\def\th{\theta}
\begin{document}
\begin{center}
\Large{\bf A Simple Continuous Parametrization of the Kasner Indices}\\
\vspace{.5cm}
\large{Alex Harvey}$^{a)}$ \\
\vspace{.15cm}
\normalsize
Visiting Scholar \\
New York University \\
New York, NY 10003 \\
\end{center} 
\vspace{1cm}
\begin{abstract}
A parametrization of the Kasner indices in terms of a continuous parameter is constructed by exploiting their representation as trilinear coordinates. This provides a clear picture of their variation through their entire range {\it vis a vis\/} each other. The parameter can be expressed as a function of time.
\end{abstract}
\section{The Kasner Metric}
The canonical form of the Kasner metric is
\be
             ds^2 = -dt^2 + t^{2a}dx^2 + t^{2b}dy^2 + t^{2c}dz^2 
\ee
where the indices $[a,\,b,\,c]$ must satisfy
\begin{subequations}\begin{align}
                   a + b + c &= 1  \label{lin}\\
            a^2 + b^2 + c^2 &= 1 \label{quad}
\end{align}\end{subequations}
This is {\em not\/} Kasner's original formulation.\footnote{An exhaustive discussion of the Kasner metric was published by Harvey \cite{harvey}.}

Equation (\ref{lin}) implies that the Kasner indices may be treated as {\it trilinear coordinates\/}  provided the {\it reference triangle\/} is equilateral.\footnote{See the Appendix.}  If so treated, then Eq. (\ref{quad}) is the locus of all points satisfying {\em both\/} Kasner conditions.
This is readily found.  If Eq. \eqref{lin} is squared and subtracted from Eq. \eqref{quad}, the result is
\be 
                         ab + bc + ca = 0
\ee
which, through division by $abc$, becomes
\be\label{circle}
              \frac{1}{a} + \frac{1}{b} + \frac{1}{b} = 0 \,.               
\ee
This is precisely the equation of a circle in trilinear coordinates if the reference triangle is, as assumed, equilateral.   

The linear Kasner condiotion implies that the altitude of the reference triangle must be equal to $1$.  The circle Eq. (\ref{circle}) will be its circumscribed circle which is thus the desired locus.  A point $P$ on the circle (see Fig. (1)) selects a set of values for $[a,\,b,\,c]$.  As P moves (clockwise) from vertex $A$ to $B$ the possible values of the indices are obtained.  As $P$ progresses further from vertex $B$ to $C$, the sets of values for $[a,\,b,\,c]$ become those for $[b,\,c,\,a]$ and similarly for $P$ moving from $C$ on to $A$.

\begin{figure}[tb]
    \centering
       \includegraphics[width=300pt,height=300pt]{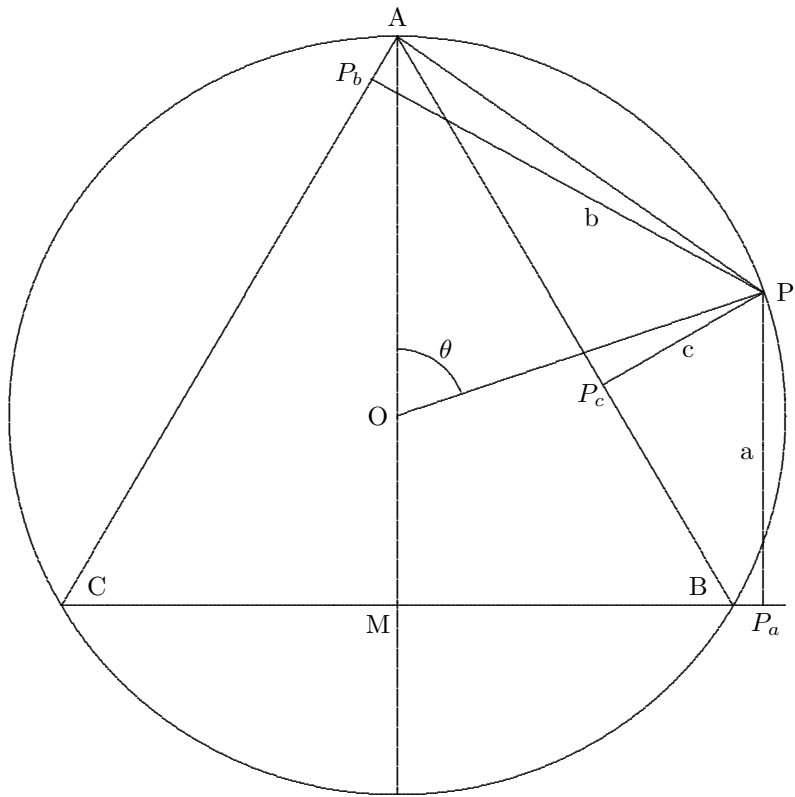}
       \caption{Kasner Coefficients in Trilinear Coordinates}
       \label{fig1}
\end{figure}

The radius of the circle is $2/3$.  The segment $OM$ is $1/3$.  Thus
\be\label{a}
                          a = \frac{1}{3} + \frac{2}{3}\cos(\th) \,.
\ee
Triangle $AOP$ is isosceles with base 
\begin{equation}                AP = 2\left(\frac{2}{3}\sin\frac{\theta}{2}\right) \,.  
\ee    
\begin{eqnarray}
                      \angle OAP &=& \angle OPA = 90^0 -\frac{\th}{2} \quad {\rm and} \nonumber \\
                    \angle P_bAP &=& 30^0 + \angle OAP \quad {\rm therefore}\nonumber \\
                    \angle P_bAP &=& 120^0 - \frac{\th}{2} 
\end{eqnarray}
Similarly,
\be
               \angle P_cAP = 60^0 - \frac{\th}{2} \,.
\ee

The altitude of the reference triangle is $1$ and the radius of the circumscribed circle is $2/3$.  Consequently 
\bea
  b &=& AP\sin \left( 120^0 - \frac{\th}{2}\right) \nonumber \\
  &=&\frac{4}{3}\sin\frac{\th}{2}\left( \frac{\sqrt{3}}{2} \cos \frac{\th}{2} + \frac{1}{2}\sin\frac{\th}{2}\right) 
\eea
and
\bea
         c &=& -AP \sin\left( 60^0 - \frac{\th}{2} \right) \nonumber\\
  &=&-\frac{4}{3}\sin\frac{\th}{2} \left( \frac{\sqrt{3}}{2} \cos\frac{\th}{2}-\frac{1}{2} \sin\frac{\th}{2}\right) 
\eea

Figure (\ref{fig2}) clearly shows how the values of the indices vary as $0 \leq \th \leq 2\pi/3$.  At the end points are the degenerate values, $[1,\, 0,\,0]$; at least one of the indices must be negative; and  midway between the endpoints are the special sets of values $[2/3,\,-1/3,\,2/3]$.  Figure (\ref{fig3}) shows the interchange of index values over a complete rotation of $OP$ over $2\pi$. 
\begin{figure}[tb]
    \centering
       \includegraphics[width=300pt,height=300pt]{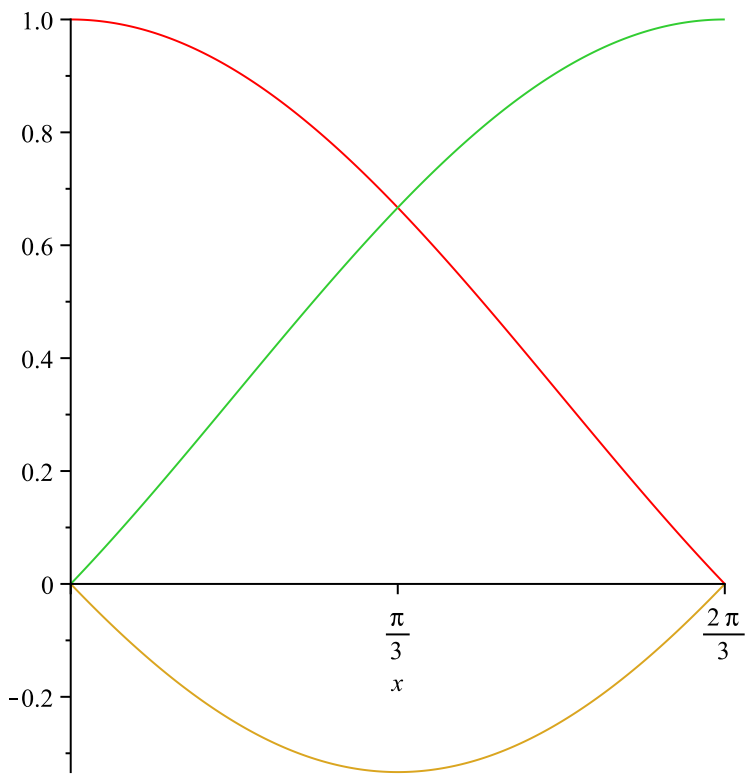}
       \caption{Numerical Range of Indices as P moves from A to B on the circle.  The indices $[a,\,b,\,c]$ are identified by red, green, and yellow respectively.}
       \label{fig2}
\end{figure}
\begin{figure}[tb]
    \centering
       \includegraphics[width=300pt,height=300pt]{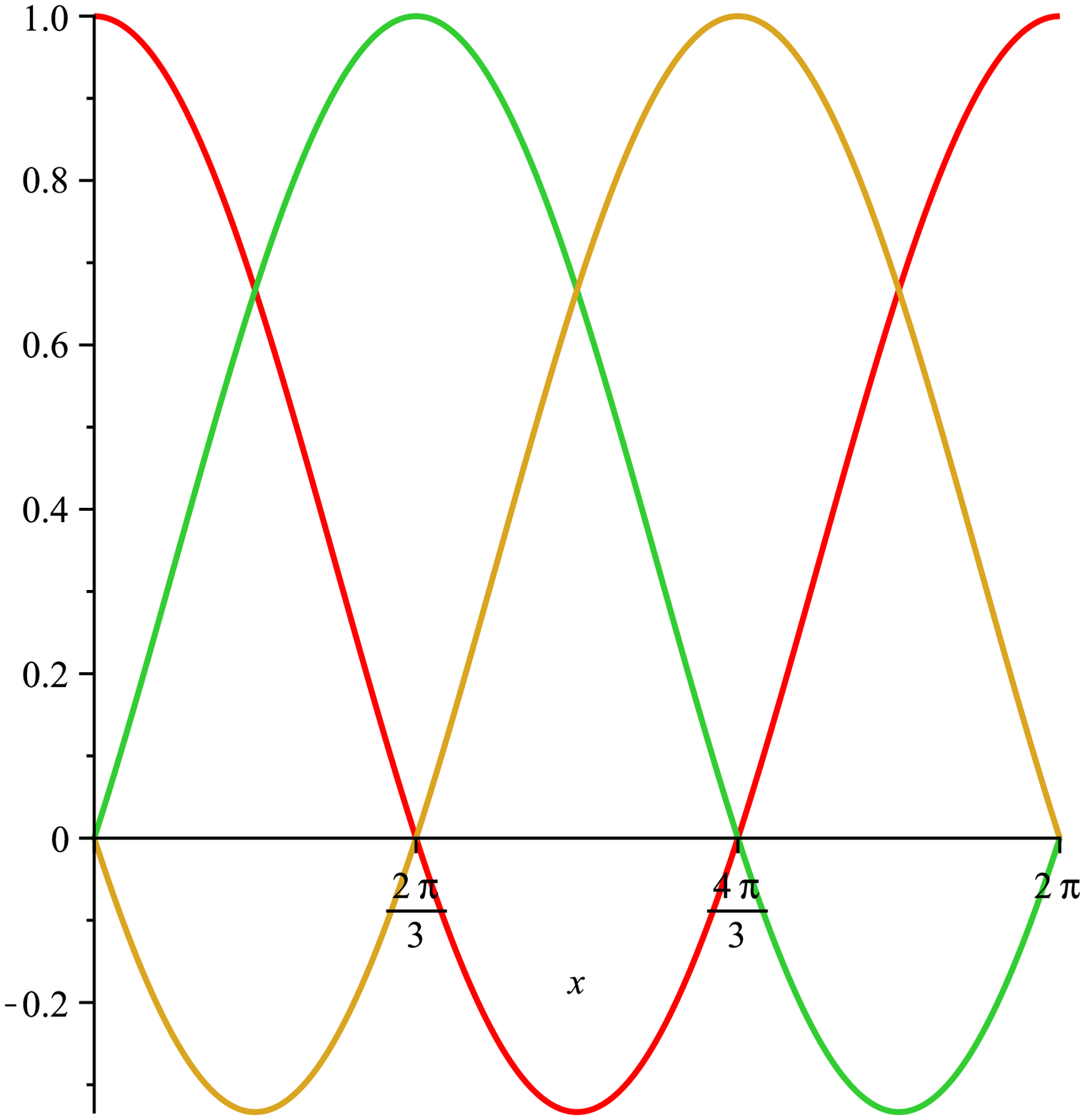}
       \caption{Complete Cycle of Indices as P moves through $2\pi$.}
       \label{fig3}
\end{figure} 
The parameter $\th$ may be made a time modulated argument, that is $\th (t)$.  It may not be amplitude modulated as this would violate the linearr Kasner condition.
\begin{figure}[tb]
      \centering
       \includegraphics[width=300pt,height=300pt]{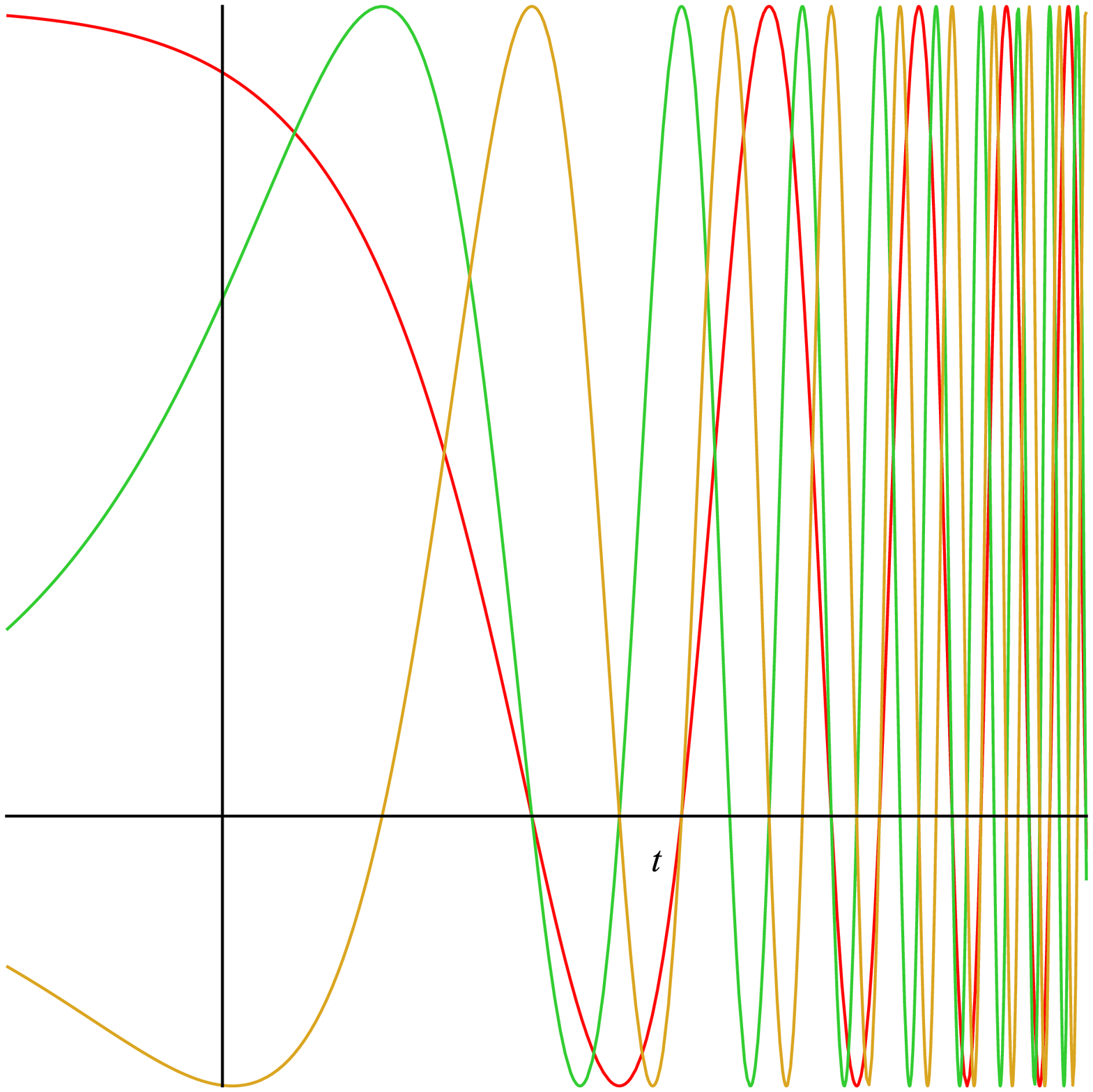}
       \label{fig4}   
      \caption{Variation of Index Values for Increasing Rate of Fluctuation}
\end{figure} 

\appendix
\newpage%
\section{Trilinear Coordinates}
Let 3 arbitrary, coplanar lines intersect to form a triangle $[A,\,B,\,C]$.\footnote{in this section, in conformity with the literature on plane geometry generally and trilinear coordinates 
\cite{ferrers} on particular, the set of symbols $[a,\,b,\,c]$ identifies the sides of the reference triangle.}  If this triangle is designated the reference triangle then the trilinear coordinates of a point, $P$, are defined by
\be\label{delta}
                        a\a+b\b+c\g = 2\Delta 
\ee
where $[\a,\,\b,\g]$ are the perpendicular distances to the respective sides.  (See Fig. (5).)  It is obvious that $\Delta$ is the area of the triangle.  The sign of each coordinate is positive or negative as the perpendiculars dropped from $P$ onto the sides of the reference triangle terminate \lq\lq inside \rq\rq or \lq\lq outside\rq\rq the (extended) sides of the triangle.  In Figure (5), the coordinates of $P_1$ are all positive.  For $P_2$, $\a_2$ and $\b_2$ are positive; $\g_2$ is negative.

If the reference triangle is equilateral with sides $s$ then Eq. (\ref{delta}) reduces to
\be
                        \a + \b + \g = \frac{2\Delta}{s} \,.
\ee
The right side may be normalized to $1$ to obtain the linear Kasner condition.

\section{Acknowledgment}
That the Kasner conditions might be interpreted in terms of trilinear coordinates was suggested by Professor Engelbert Schucking.   The diagrams were prepared by Dr. Eugene Surowitz.  His work is deeply appreciated.
\begin{figure}[tb]
    \centering
       \includegraphics[width=300pt,height=300pt]{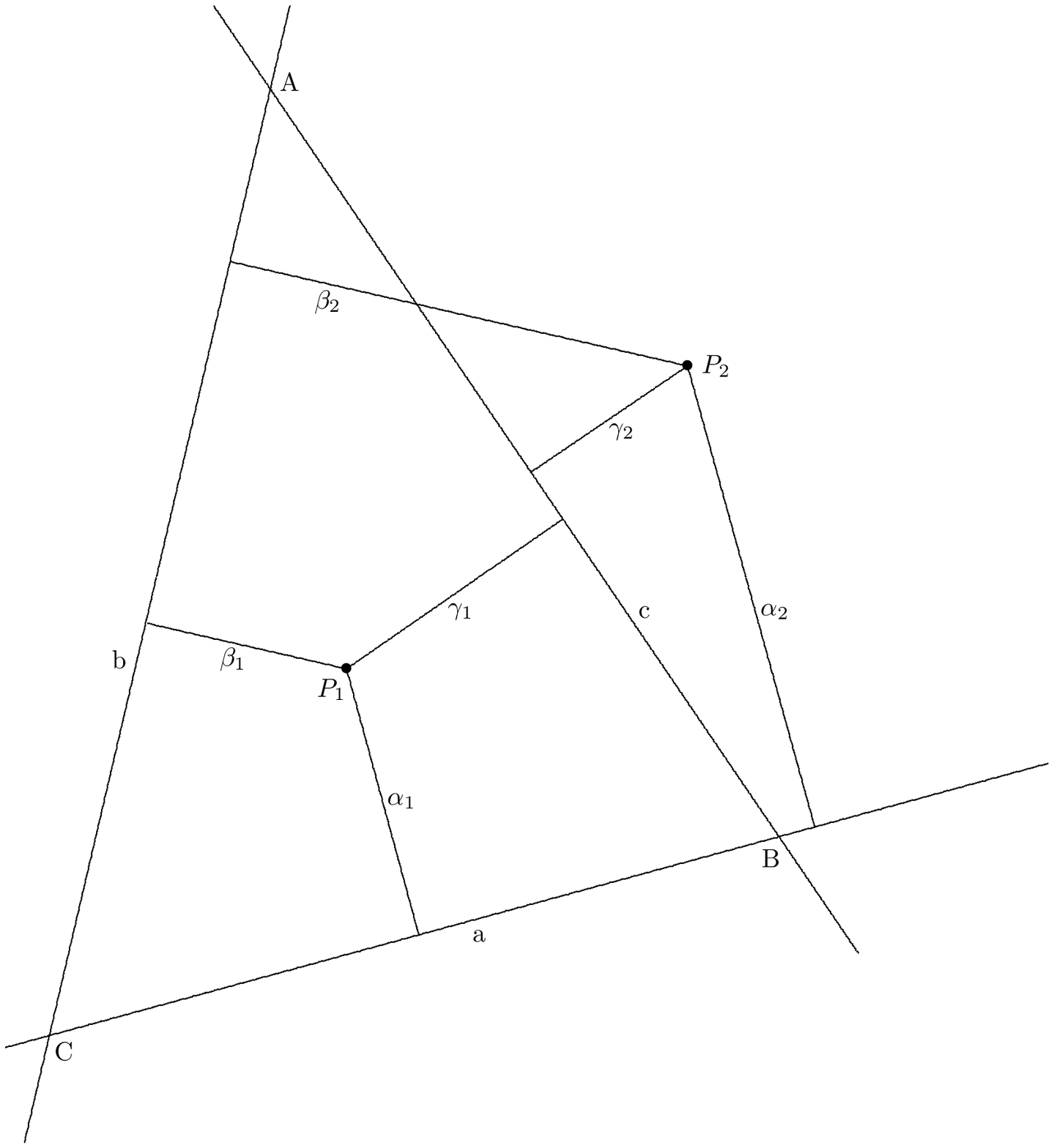}
           \caption{Arbitrary Reference Triangle}
      \label{fig5}
\end{figure}\\
\vspace{3mm}
${}^{a)}$Prof. Emeritus, Queens College, City University of New York, email: ah30@nyu.edu


\end{document}